\documentclass[useAMS,usenatbib]{mn2e}
\usepackage{graphicx}
\usepackage{lscape}

\title[High precision parameters of BY Dra]
{New high precision orbital and physical parameters of the
double-lined low-mass spectroscopic binary BY Draconis}
\author[K. G. He{\l}miniak et al.]
{K. G. He{\l}miniak$^{1,2}$\thanks{E-mail: xysiek@ncac.torun.pl},
M. Konacki$^{1,3}$, M. W. Muterspaugh$^{4,5}$, S. E. Browne$^{6}$,
\newauthor
A. W. Howard$^{7}$ and S. R. Kulkarni$^{8}$
\\
$^{1}$Nicolaus Copernicus Astronomical Center, 
Department of Astrophysics, ul. Rabia\'{n}ska 8 , 
87-100 Toru\'{n}, Poland\\
$^{2}$Departamento de Astronom\'{i}a y Astrof\'{i}sica, Pontificia Universidad
Cat\'{o}lica, Casilla 306, Santiago, Chile\\
$^{3}$Astronomical Observatory, A. Mickiewicz University, ul. S{\l}oneczna 36, 
60-286 Pozna\'{n}, Poland\\
$^{4}$Department of Mathematics and Physics, College of Arts
and Sciences, Tennessee State University, Boswell Science Hall,
Nashville, \\TN 37209, USA\\
$^{5}$Tennessee State University, Center of Excellence in Information
Systems, 3500 John A. Merritt Blvd., Box No. 9501,
Nashville, \\TN 37203-3401, USA\\
$^{6}$Experimental Astrophysics Group, Space Sciences Laboratory, University of California, 7 Gauss Way, Berkeley, CA 94720 \\
$^{7}$UC Berkeley Astronomy Dept., Campbell Hall MC 3411, 
Berkeley, CA 94720, USA\\
$^{8}$Division of Physics, Mathematics and Astronomy, California
Institute of Technology, Pasadena, CA 91125, USA
}
\begin{document}

\date{Accepted ... Received ...; in original form ...}

\pagerange{\pageref{firstpage}--\pageref{lastpage}} \pubyear{2011}

\maketitle

\label{firstpage}

\begin{abstract}
We present the most precise to date orbital and physical parameters
of the well known short period ($P=5.975$ d), eccentric ($e=0.3$)
double-lined spectroscopic binary BY Draconis,
a prototype of a class of late-type, active, spotted flare stars.
We calculate the full spectroscopic/astrometric orbital solution by 
combining our precise radial velocities (RVs) and the archival astrometric
measurements from the Palomar Testbed Interferometer (PTI). The RVs 
were derived based on the high resolution echelle spectra taken between
2004 and 2008 with the Keck~I/HIRES, Shane/CAT/HamSpec and TNG/SARG
telescopes/spectrographs using our novel iodine-cell technique for
double-lined binary stars. The RVs and available PTI astrometric data
spanning over 8 years allow us to reach 0.2-0.5\% level of 
precision in $M \sin^3 i$ and the parallax but the geometry of the 
orbit ($i\simeq154^\circ$) hampers the
absolute mass precision to 3.3\%, which is still an order of magnitude 
better than for previous studies. We compare our results with a set of
Yonsei-Yale theoretical stellar isochrones and conclude that BY Dra is probably
a main sequence system more metal-rich than the Sun. Using the orbital 
inclination and the available rotational velocities of the components, 
we also conclude that the rotational axes of the components are likely
misaligned with the orbital angular momentum. Given BY Dra's main sequence 
status, late spectral type and the relatively short orbital period,
its high orbital eccentricity and probable spin-orbit misalignment
are not in agreement with the tidal theory. This disagreement
may possibly be explained by smaller rotational velocities of the 
components and the presence of a substellar mass
companion to BY Dra AB.
\end{abstract}


\section{Introduction}

Regular studies of BY Dra (Gl 719, HD 234677, HIP 91009, 
NLTT 46684, BD+51 2402) started in mid 40's when M\"{u}nch 
noted the calcium H and K lines to be in emission \citep{mun44}. 
This fact was later confirmed by \citet{pop53} who also noted 
strong emission Balmer series lines. In one of his spectrograms the 
emission was particularly strong which led to a conclusion that BY Dra 
may be a member of a new group of flare stars \citep{pop53}. Photometric 
monitoring was then carried out \citep[e.g.][]{mas55} but the variability 
was not confirmed until 1966 when Chugainov obtained a quasi-sinusoidal 
light curve with an amplitude of 0.23 mag and a 
period of 3.826 d \citep{chu66}, later interpreted as a 
rotation of a spotted star \citep{krz69}. No flares were then observed. The first 
photometric flares were reported by \citet{cri68} who later observed
twelve flares that occurred between July 1967 and July 1970 
\citep{cri70,cri71}. \cite{krz69} confirmed the sinusoidal 
variability with a $\sim$3.826 d period and noted the variation 
of its amplitude. Many subsequent studies of BY Dra's
variability have been carried out and the most up-to-date value of 
the rotational period $P_{rot} = 3.8285$ d is
given by \citet{pet92} as an average period for their entire 
1965-1989 data set.

The double-lined spectroscopic nature was revealed by \citet{krz67}. 
They announced a period of 5.981~d but never published their full 
orbital solution. It was done later by \citet{bop73} on the basis 
of 23 spectra taken between June 1966 and July 1971 at the Hale 
and McDonald observatories, 15 of which showed unblended 
Ca II H and K lines. \citet{bop73} also performed an analysis of the spots 
on the surface of BY Dra and estimated the rotational velocity of the
primary (spotted) component to be $\sim$5 km s$^{-1}$ and the 
(rotational) inclination to be $\sim$30 deg. Since then 
BY Dra became a prototype of a new class of stars characterized by 
a late type, brightness variation caused by spots, rapid rotation and
strong emission in H and K lines. The short orbital period also
seems to be a characteristic for most of BY Dra-type stars 
\citep{bop80}.

The orbital solution was later improved by \citet{vog79} on the basis 
of high-resolution reticon spectra. \citet{vog79} also found the 
projected rotational velocity of the primary to be 8.5 km s$^{-1}$
under the assumption of the rotational inclination being the same as
the orbital one (spin-orbit alignment). They estimated the radius
of the primary to be greater than 0.9 R$_{\odot}$ which led to a conclusion that
BY Dra was a pre-main-sequence system. This conclusion was supported by 
the large brightness ratio despite the mass ratio being close to 1
($q = 0.98$), the Barnes-Evans visual surface brightness relation 
\citep{bar78} and the inequality of the rotational and orbital periods. 
However, the assumption of the spin-orbit alignment in close binary 
systems was later criticized in several works, e.g. \citet{gle95}. 
The orbital parameters as well as the value of the projected rotational 
velocities for both components were shortly after improved by \citet{luc80}. 
They used new measurements from CORAVEL and obtained rotational velocities 
$v_1 \sin i = 8.05$ and $v_2 \sin i = 7.42$ km s$^{-1}$ and the mass 
ratio $q$ = 0.89 significantly more different from 1 than that of \citet{vog79}.
They also estimated the magnitude difference (1.15 mag) and the primary's
radius (1.2 - 1.4 R$_{\odot}$) but noted that a higher macroturbulence 
velocity would reduce the radius estimation by a factor of 2.

The most up-to-date spectroscopic orbital solution was given together
with the first astrometric solution by \citet{bod01}. They combined
the archival RV measurements with the visibility based ($V^2$) astrometric
measurements obtained with the Palomar Testbed Interferometer \citep[PTI;][]{col99} 
in 1999. Their orbital inclination (152 deg, retrograde orbit) agrees 
with the first estimations of the rotational inclination \citep{bop73} 
but not with the later ones \citep{gle95}. 

Finally it is worth noting that BY Dra is a hierarchical 
multiple system. A common proper-motion companion was found by 
\citet{zuc97} about 16.7 arc-sec to the northeast of the primary. 
From the visual and infrared photometry of BY Dra C they also 
deduced that this component is a normal M5 dwarf at least 
$3\times 10^8$ yr old which makes the pre-main-sequence 
nature of BY Dra less probable. Yet another putative component 
is reported in the \emph{Hipparcos} Double 
and Multiple System Annex \citep{esa97}. A photocentric circular 
orbital solution with a period of 114 d and 113 deg inclination
is reported. \citet{bod01} however demonstrated that this is an 
improbable solution since the 4-th body would produce significant 
perturbations to the BY Dra AB radial velocities but no such periodicity 
is seen in the archival RVs. 

We spectroscopically observed BY Dra over the years 2004-2008
using a combination of high resolution echelle spectrographs
HIRES (10-m Keck I), SARG (3.5-m TNG) and HamSpec (3-m Shane 
telescope) as a part of our ongoing RV search for circumbinary 
planets \citep{kon09a,kon10}. Even though we knew that BY Dra
was too variable to allow us reach an RV precision sufficient to 
detect planets, it was nevertheless observed to make use
of an extensive and publicly available set of PTI $V^2$ 
measurements spanning now over 8 years.

In this paper we present a new orbital solution and the
orbital and physical parameters of the BY~Dra~AB binary, derived 
with a precision of over an order of magnitude better than by 
\citet{bod01}. Thanks to our superior iodine-cell-based radial velocities
and the full set of PTI visibilities we are able to put strong
constraints on the nature of the system. In Sections 2 and 3 we 
present the data --- $V^2$s and RVs. In Section 4 we describe their 
modeling. The results of our data modeling are presented in 
Section 5 and the state of BY Dra is then discussed in Section 6.

\section{Visibilities}

\begin{table*}
\caption{Absolute values of the radial velocities of BY Dra with their 
errors and the best-fit $O-C$s. The formal error is denoted with $\sigma$ and the
adopted final error with $\epsilon$. The subscript "1" is for the primary
and "2" for the secondary. K/H denotes the measurements from 
Keck~1/HIRES, T/S from TNG/SARG and S/H from Shane/HamSpec.}
\label{tab_allrv}
\begin{tabular}{cccccccccc}
\hline \hline
TDB - 2400000 & $v_1$ & $\sigma_1$ & $\epsilon_1$ & $O-C_1$ & $v_2$ & $\sigma_2$ & $\epsilon_2$ & $O-C_2$ & Tel./Spec. \\
& (km s$^{-1}$) & (km s$^{-1}$) & (km s$^{-1}$) & (km s$^{-1}$) & (km s$^{-1}$) & (km s$^{-1}$) & (km s$^{-1}$) & (km s$^{-1}$) & \\
\hline
53276.214636 &  -49.47297 &  0.00668  &  0.15015  &   0.05236 &    1.74203 &  0.01162  &  0.15045  &  -0.04117 & K/H \\ 
53276.218251 &  -49.29837 &  0.00822  &  0.15022  &   0.06779 &    1.57802 &  0.01371  &  0.15063  &  -0.02414 & K/H \\ 
53276.274388 &  -46.62269 &  0.00994  &  0.15033  &   0.14884 &   -1.24054 &  0.01193  &  0.15047  &   0.10849 & K/H \\ 
53276.277251 &  -46.46912 &  0.01018  &  0.15034  &   0.16444 &   -1.39294 &  0.01152  &  0.15044  &   0.11303 & K/H \\ 
53276.371688 &  -41.65463 &  0.00990  &  0.15033  &   0.19386 &   -6.80759 &  0.01758  &  0.15103  &   0.14139 & K/H \\ 
53276.381066 &  -41.16056 &  0.00934  &  0.15029  &   0.19369 &   -7.39337 &  0.02663  &  0.15235  &   0.11782 & K/H \\ 
53276.383829 &  -41.01684 &  0.01020  &  0.15035  &   0.19138 &   -7.57152 &  0.02920  &  0.15282  &   0.10579 & K/H \\ 
53328.260862 &  -34.23894 &  0.01017  &  0.15034  &  -0.13033 &  -15.80064 &  0.01163  &  0.15045  &  -0.00187 & K/H \\ 
53329.192929 &  -55.69433 &  0.00958  &  0.15031  &  -0.36569 &    8.16117 &  0.01185  &  0.15047  &  -0.19837 & K/H \\ 
53567.417144 &  -37.51257 &  0.00765  &  0.15019  &  -0.07442 &  -12.24447 &  0.01409  &  0.15066  &  -0.23557 & K/H \\ 
53654.287570 &   -2.81022 &  0.00707  &  0.15017  &  -0.26482 &  -51.63684 &  0.01257  &  0.15053  &   0.04275 & K/H \\ 
53655.270877 &   -8.62844 &  0.00867  &  0.15025  &  -0.20946 &  -44.96407 &  0.00776  &  0.15020  &   0.05011 & K/H \\ 
53656.250010 &  -21.53447 &  0.01434  &  0.15068  &   0.29488 &  -29.89713 &  0.02974  &  0.15292  &  -0.12965 & K/H \\ 
54191.188978 &  -13.36607 &  0.02168  &  0.15156  &   0.10101 &  -39.92790 &  0.03336  &  0.15366  &  -0.10395 & T/S \\ 
54192.164756 &   -2.95488 &  0.01336  &  0.15059  &  -0.10836 &  -52.02801 &  0.03054  &  0.15308  &  -0.10339 & T/S \\ 
54247.147226 &  -12.62161 &  0.02067  &  0.15142  &  -0.03078 &  -40.66218 &  0.02748  &  0.15250  &   0.19585 & T/S \\ 
54275.086679 &   -7.54149 &  0.01298  &  0.15056  &   0.03906 &  -46.51656 &  0.02529  &  0.15212  &   0.00909 & T/S \\ 
54281.357833 &   -3.42439 &  0.02136  &  0.15151  &   0.18914 &  -50.40787 &  0.02344  &  0.15182  &   0.02977 & S/H \\ 
54290.431083 &  -38.18861 &  0.00594  &  0.15012  &  -0.18164 &  -11.42422 &  0.00905  &  0.15027  &  -0.06282 & K/H \\ 
54290.596520 &  -41.90239 &  0.00805  &  0.15022  &  -0.07812 &   -7.00027 &  0.01280  &  0.15055  &   0.01660 & K/H \\ 
54727.249888 &  -52.87236 &  0.01649  &  0.15090  &  -0.09239 &    5.11252 &  0.02114  &  0.15148  &  -0.36239 & S/H \\ 
54728.248858 &  -45.72092 &  0.02849  &  0.15268  &  -0.09889 &   -2.56209 &  0.02087  &  0.15145  &   0.07559 & S/H \\ 
54752.198503 &  -43.21949 &  0.01716  &  0.15098  &  -0.08641 &   -5.07445 &  0.01632  &  0.15089  &   0.39454 & S/H \\ 
54789.116481 &   -4.71313 &  0.02199  &  0.15160  &   0.08852 &  -49.23474 &  0.04418  &  0.15637  &  -0.14975 & S/H \\ 
\hline
\end{tabular}
\end{table*}

Often the main observable in the interferometric observations at 
optical or infrared wavelength is the normalized amplitude of the 
coherence function --
a fringe pattern contrast, commonly known as the visibility 
(squared, V$^2$) of the  interferometric fringes, calculated by 
definition as follows \citep{bod99}:
\begin{equation}
V^2 = \left(\frac{I_{max} - I_{min}}{I_{max} + I_{min}}\right)^2
\end{equation}
where $I_{max}$ and $I_{min}$ are the maximum and minimum intensity of the 
fringe pattern respectively. For a given object the observed $V^2$ depends on 
its morphology and the projected baseline vector of a two-aperture interferometer 
$\bmath{B_\bot}$ onto a plane tangent to the sky. For binaries, $V^2$ varies
also due to the orbital motion of the components. In the case of a binary, 
approximated by two uniform disks, the squared visibility can be modeled 
as follows \citep[see e.g.][]{bod99}:
\begin{equation}
V^2_{binary} = \frac{V^2_1 + r^2 V^2_2 + 
2rV_1V_2\cos(2\pi\bmath{B_\bot}\cdot\bmath{\Delta s}/\lambda)}{(1+r)^2}
\end{equation}
where $V_{1,2}$ are the visibilities of uniform disks (components) of 
the angular diameters $\theta_1$ and $\theta_1$ and are calculated as
follows:
\begin{equation}
\label{eq_v}
V_i^2 = \left ( \frac{2J_1(\pi\theta_i B_\bot \lambda)}
{\pi\theta B_\bot \lambda} \right)^2
; i = 1,2; B_\bot = ||\bmath{B_\bot}||
\end{equation}
where $r$ is the brightness ratio at the observing wavelength $\lambda$,
$J_1(x)$ is the first order Bessel function 
and $\bmath{\Delta s} = (\Delta \alpha, \Delta \delta)$ is the 
separation vector between the primary and the secondary in the plane 
tangent to the sky. This vector is related to the Keplerian orbital 
elements, orbital period $P$, eccentric anomaly $E$ (from the Kepler 
equation $E - e \sin E = M$), and the parallax $\kappa$ in the usual
way \citep{kam67}.

A visibility measurement needs to be calibrated by observing at 
least one calibration source before or after a target observation.
The calibrator is typically a single star with a known diameter 
and its visibility $V^2_{cal}$ is given by Relation \ref{eq_v}.
The correction factor $f$ which should be applied to the observed 
target $V^2$ is simply the ratio $f = V^2_{cal}/V^2_{cal-meas}$ where
$V^2_{cal-meas}$ is the measured calibrator visibility. The ``true''
target visibility is then
\begin{equation}
V^2_{true} = f V^2_{measured}.
\end{equation}

Uncalibrated visibilities of BY Dra were extracted from the NASA 
Exoplanet Science Institute (NExSci) database of the PTI 
measurements\footnote{https://nexsciweb.ipac.caltech.edu/pti-archive/secure/main.jsp}.
These measurements were made in $K$ (2.2 $\mu$m) and $H$ (1.6 $\mu$m) 
bands. They were calibrated using the standard tools provided by NExSci 
({\tt getCal} and {\tt wbCalib}). As the calibration objects we used 
HD~177196 (A7V, $V=5.0$ mag, $K=4.5$ mag, diameter $\theta=0.42$ mas, 
6.6 deg from BY Dra) and HD~185395 
(F4V, $V=4.5$ mag, $K=3.5$ mag, $\theta=0.73$ mas, 9.9 deg) as in \cite{bod01}. 
We do not list these measurements as they can be easily obtained using 
the NExSci database and tools.

\section{Radial velocities}

Our high-resolution echelle spectra of BY Dra were obtained during 17
nights between September 2004 and November 2008. We collected 24 spectra
using Keck~I/HIRES (K/H, 15 spectra), TNG/SARG (T/S, 4) and 
Shane/HamSpec (S/H, 5) telescopes/spectrographs. Our spectra have the
resolutions R $\sim$ 67 000 for K/H, 86 000 for T/S and 60 000 for S/H. 
The typical signal to noise ratio ($SNR$) per collapsed pixel at 550 nm was 
$\sim$250 for K/H, $\sim$90 for T/S and $\sim$60 for S/H.
The basic reduction (bias, dark, flatfield, scattered light subtraction) was
done with the \textsc{ccdred} and \textsc{echelle} packages from 
\textsc{iraf}\footnote{\textsc{iraf} is written and supported by the 
\textsc{iraf} programming group at the National Optical Astronomy 
Observatories (NOAO) in Tucson, AZ. NOAO is operated by the Association of 
Universities for Research in Astronomy (AURA), Inc. under cooperative 
agreement with the National Science Foundation. http://iraf.noao.edu/}. 
The wavelength solution and radial velocities were obtained with our 
novel procedure based on the iodine cell technique \citep{kon09,kon09a,kon10}.
This procedure employs a tomographic disentangling of the component 
spectra of double-lined spectroscopic binaries (SB2s) implemented 
through a maximum entropy method and the two-dimensional cross-correlation 
technique \textsc{todcor} \citep{zuc94} using synthetic spectra derived with 
\textsc{atlas~9} and \textsc{atlas~12} codes \citep{kur95} as templates 
for the first approximation of the RVs. With this 
approach it is possible to reach up to 2 m s$^{-1}$ precision in RVs 
for components of SB2s \citep{kon09a} but in the case of BY Dra 
the precision is hampered by the activity of the star (presence of 
spots) and the relatively rapid rotation of both components.

In Table \ref{tab_allrv} we list our RV measurements 
together with their uncertainties and the best-fit $O-C$s. 
The formal errors, $\sigma$, were calculated from the scatter between 
orders and predominantly reflect a high SNR of our spectra. The 
formal errors underestimate the true RV scatter (due to activity) and 
the resulting reduced $\chi^2$ of the spectroscopic orbital fit was much larger 
than 1. Hence to obtain a conservative estimation of the parameters' 
errors (and the reduced $\chi^2$ close to 1) we added in quadrature a 
systematic error $\sigma_{sys}$ of 150 m s$^{-1}$. 
Let us note that spots can easily induce RV variations at the level
of a few hundreds of m s$^{-1}$ so the RV variability of BY Dra
is not surprising \citep[see e.g.][]{hel11a,hel11b}.
We also had to adopt 
small shifts between each data set as is explained in \cite{kon10}. 
The best fit values of the shifts can be found in Table \ref{tab_orb} 
in Section \ref{sec_res}. We do not include the CORAVEL data \citep[from][]{luc80}
since their precision is substantially worse than ours.

\section{Modeling}

\begin{figure}
\includegraphics[width=\columnwidth]{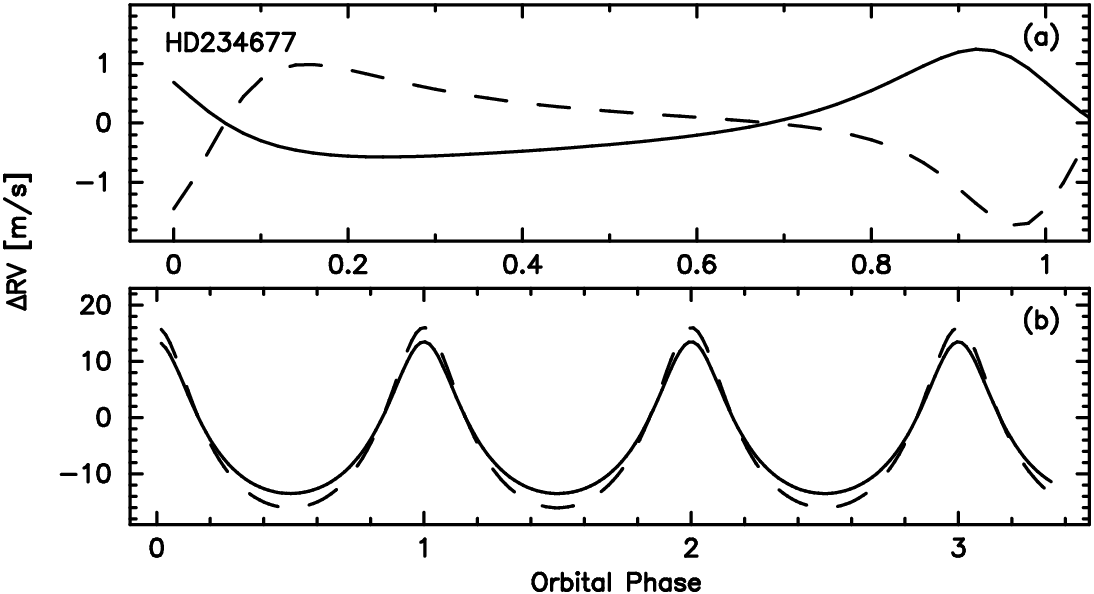}
\caption{Radial velocity variations of the primary (solid lines) 
and secondary (dashed lines) stars of BY Dra (HD234677)
as a function of the orbital phase. The top panel (a) shows
the RV variations due to the tidal distortion of the stars
and the bottom panel (b) due to the combined 
gravitational redshift and transverse Doppler effects
which together are the dominant term of the relativistic correction.}
\label{tid_GR}
\end{figure}

We combined all $V^2$ and RV measurements in a simultaneous least-squares
fit to derive the full orbital solution and the physical parameters
of BY Dra. We used our own procedure which minimizes the $\chi^2$
function with a least-squares Levenberg-Marquardt algorithm. The
procedure fits a Keplerian orbit with corrections to the RVs
due to tidal distortions of the components and relativistic effects. 
In order to model the tidal term we use the Wilson-Devinney 
(WD) code \citep{wil71} as is explained in \cite{kon10} and 
assume several parameters of BY Dra listed in Table \ref{tab_ass}.
Note that both the relativistic and tidal effects are much smaller
than the RV scatter (see Fig.~\ref{tid_GR}) but we decided to include them in the
RV model anyway to maintain a consistent treatment of our iodine cell 
based RVs as in \cite{kon10}. Apparent stellar diameters were assumed
to agree with the estimates of the radii from Section 6.2, 
since the components are too small and act like point sources. 

For a combined $V^2$+RV solution our software evaluates the
the period $P$, standard Keplerian elements: major semi-axis $\hat{a}$ 
(of B relatively to A -- apparent astrometric in mas), 
inclination $i$, eccentricity $e$, longitude of pericenter 
$\omega$, longitude of ascending node $\Omega$, time of 
periastron passage $T_p$; velocity amplitudes $K_1$ and $K_2$,
systemic velocity $v_0$, flux ratios in the observing bands $r_H$ 
and $r_K$, and a set of shifts in radial velocities between the
two components as well as between the data sets from each 
telescope/spectrograph. On this basis the software calculates
such absolute physical parameters like the absolute major
semi-axis $a_{1,2}$ (relatively to the baricentre -- in AU), 
absolute components' masses $M_1$ and $M_2$, 
magnitude differences $\Delta H$ and $\Delta K$, and 
parallax $\kappa$. The uncertainty of every parameter
is a combination of formal best-fit least-squares errors and 
systematic errors as is explained in \cite{kon10}. For the
systematic errors we assumed the following estimates for 
additional uncertainties related to the $V^2$ data reduction
(1) 0.01 percent in the baseline vector coordinates, 
(2) 0.5 percent in $\lambda$ and (3) 10 percent in the calibrator 
and binary components diameters. For the RVs we assumed (4) 10 
percent in all the parameters from Table 2 except for the temperatures 
for which we assumed an uncertainty of 2 percent and for the
metallicities we assumed an uncertainty of 0.05 dex. 

\begin{figure}
\includegraphics[width=\columnwidth]{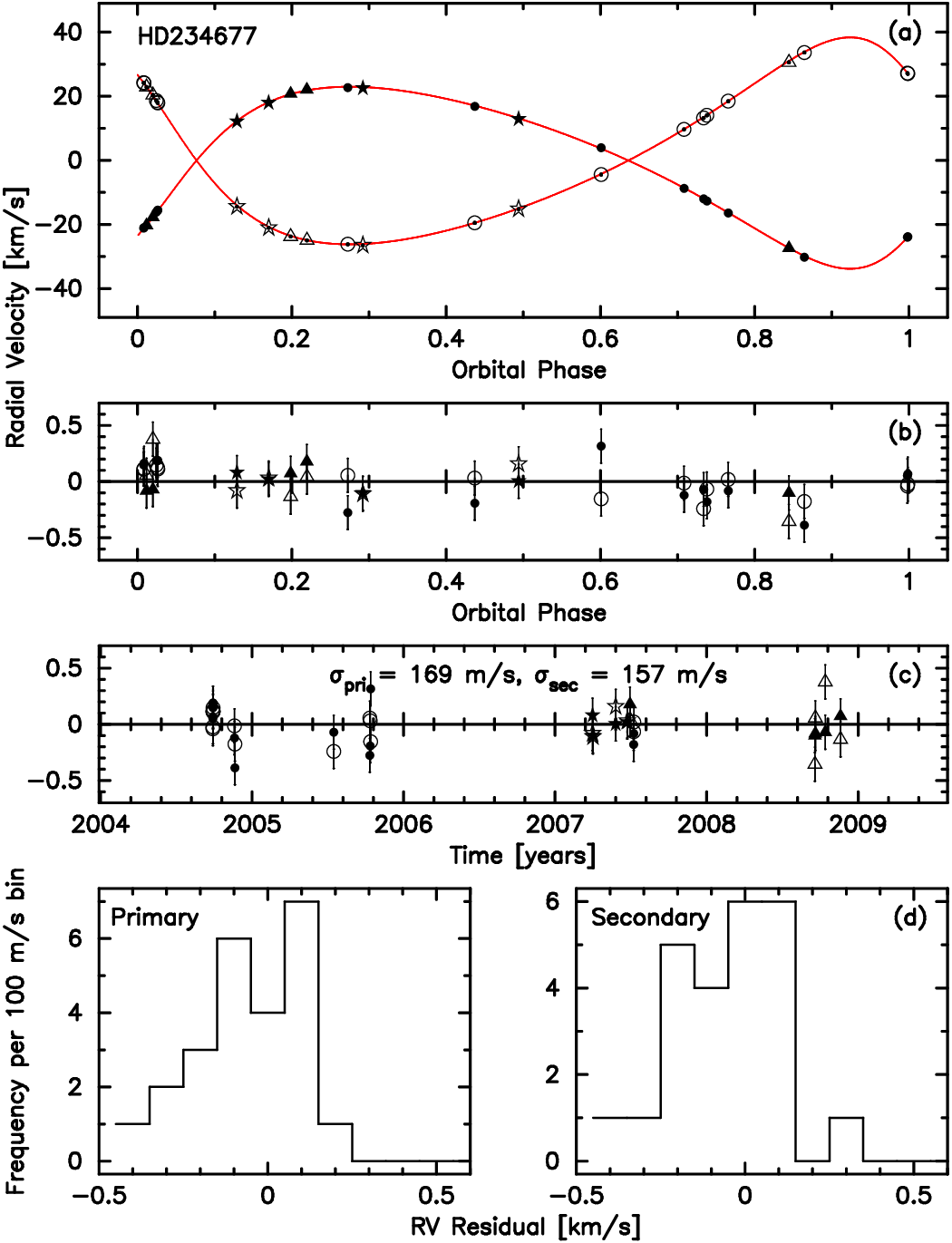}
\caption{Observed and modeled radial velocities of BY Dra as a
function of the orbital phase (a), and their best-fit residuals as
a function of the orbital phase (b) and time (c). The histograms
of the residuals for the primary and secondary (d). The Keck I/HIRES
measurements are
denoted with circles, Shane/CAT/HamSpec with triangles and TNG/SARG 
with stars. Colour version of the Figure is available 
in the on-line version of the article.}\label{rvs}
\end{figure}

\begin{figure}
\includegraphics[width=\columnwidth]{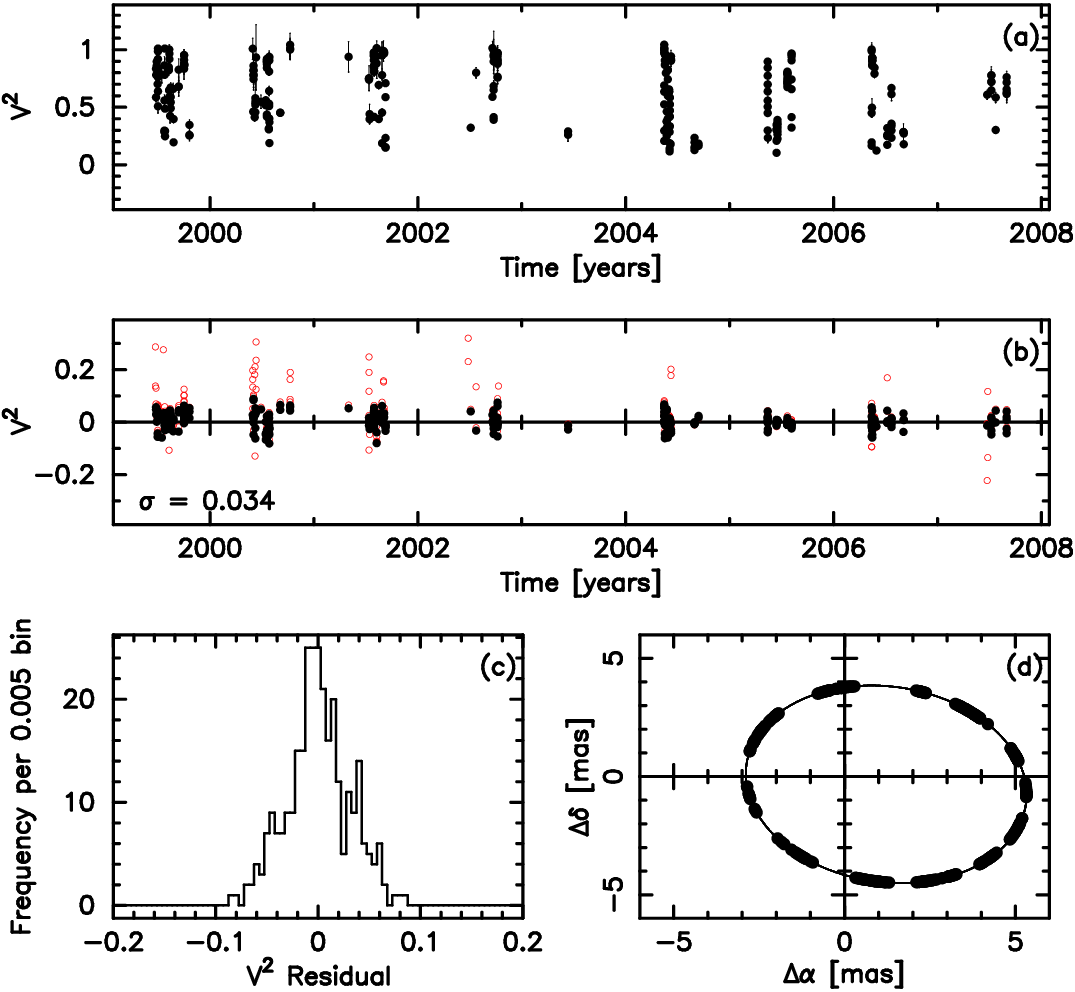}
\caption{Visibility measurements of BY Dra as a function of time (a),
and their best-fit residuals as a function of time (b) and histogram (c).
The 299 measurements used to determine the best-fit orbital solution
are denoted with black filled circles. The corresponding orbital coverage 
and the relative orbit is shown in the panel (d). Colour version of 
the Figure is available in the on-line version of the article.}\label{vis}
\end{figure}

\section{Results}\label{sec_res}

\begin{table}
\caption{Assumed parameters of BY Dra}
\label{tab_ass}
\begin{tabular}{lcc}
\hline \hline
Parameter & Primary & Secondary \\
\hline
Effective temperature, $T$ (K) & 4000 & 4000 \\
Potential, $\hat{\Omega}$ &  24.0 & 24.5 \\
Synchronization factor, $F$ & 1.95 & 1.95 \\
Gravity darkening exponent, $g$ & 0.3 & 0.3 \\
Albedo, $A$ & 0.5 & 0.5 \\
Apparent diameter, $\theta$ (mas) & 0.6 & 0.5 \\
Metallicity & \multicolumn{2}{c}{0.0} \\
\hline
\end{tabular}
\end{table}

\begin{table}
\caption{Best-fitting orbital solution and its parameters for BY Dra.}
\label{tab_orb}
\begin{tabular}{lr}
\hline \hline
Parameter & Value($\pm$) \\
\hline
\multicolumn{2}{c}{\emph{Orbital solution}}\\
Apparent major semi-axis, $\hat{a}$ (mas) & 4.4472(91) \\
Period, $P$ (d) & 5.9751130(46) \\
Time of periastron, $T_p$ (TDB-2450000.5) & 3999.2144(21) \\
Eccentricity, $e$ & 0.30014(62) \\
Longitude of periastron, $\omega$ (deg) & 230.33(17) \\
Longitude of ascending node, $\Omega$ (deg) & 152.30(10) \\
Inclination, $i$ (deg) & 154.41(29) \\
Magnitude difference in K band, $\Delta K$ (mag) & 0.530(11) \\
Magnitude difference in H band, $\Delta H$ (mag) & 0.60(23) \\
Velocity amplitude, primary, $K_1$ (km s$^{-1}$) & 28.394(60) \\
Velocity amplitude, secondary, $K_2$ (km s$^{-1}$) & 32.284(61) \\
Mass ratio, $q$	& 0.8795(25) \\
Gamma velocity, $v_0$ (km s$^{-1}$) & -25.484(46) \\
 &  \\
\multicolumn{2}{c}{\emph{Velocity offsets (all in km s$^{-1}$)}}\\
Secondary vs primary & -0.088(67) \\
SARG vs HIRES, primary & -0.216(104) \\
SARG vs HIRES, secondary & -0.343(105) \\
HamSpec vs HIRES, primary & 0.076(83) \\
HamSpec vs HIRES, secondary & -0.067(85) \\
 &  \\
\multicolumn{2}{c}{\emph{Least-squares fit parameters}}\\
Number of RV measurements, total & 48 \\
Number of RV measurements, HIRES &  30\\
Number of RV measurements, SARG &  8\\
Number of RV measurements, HamSpec &  10\\
Number of V$^2$ measurements & 299 \\
Combined RV rms, prim./sec. (km s$^{-1}$) & 0.169/0.157 \\
Visibilities $V^2$ rms & 0.0312 \\
RV $\chi^2$, primary/secondary & 29.17/24.65 \\
Visibilities $V^2$ $\chi^2$ & 395.3 \\
Degrees of freedom, $DOF$ & 330 \\
Total reduced $\chi^2$, $\chi^2/DOF$ & 1.361 \\
\hline
\end{tabular}
\end{table}

\begin{table}
\caption{Physical parameters of BY Dra}
\label{tab_phys}
\begin{tabular}{lcc}
\hline \hline
Parameter & Primary & Secondary \\
\hline
Major semi-axis, $a$ (10$^{-2}$ AU) & 3.4437(73) & 3.9155(74) \\
Major semi-axis, $a$ (R$_\odot$) & 7.400(16) & 8.414(16) \\
$M$ sin$^3 i$ (M$_\odot$) & 0.06387(28) & 0.05618(26) \\
Mass, $M$ (M$_\odot$) & 0.792(26) & 0.697(23) \\
M$_{K,2MASS}$ (mag) & 4.269(21) & 4.799(22) \\
M$_{H,2MASS}$ (mag) & 4.420(86) & 5.020(149) \\
Parallax, $\kappa$ (mas) & \multicolumn{2}{c}{60.43(12)} \\
Distance, $d$ (pc) & \multicolumn{2}{c}{16.548(35)} \\
\hline
\end{tabular}
\end{table}

The results of our modeling are collected in Tables \ref{tab_orb} and
\ref{tab_phys}. Figure \ref{rvs} shows our RVs together
with the best fitting orbital solution and the corresponding residuals
and their histograms. Figure \ref{vis} shows the same for 
the PTI $V^2$ measurements. The resulting
astrometric orbit of component B relative to A is shown in
the panel (d). In Table \ref{tab_orb} we show the orbital parameters 
for BY Dra, the velocity offsets and other parameters related to the 
quality of the fit. The absolute physical parameters are listed in 
Table \ref{tab_phys}.

As one can see, we were able to reach $\sim$0.2 \% of precision in
velocity amplitudes, despite such obstacles like the presence of spots 
or some rotational broadening of spectral lines. This level of quality has 
direct implication for the precision of mass ratio $q$ (0.28 \%), 
$M\sin^3 i$ (0.44 and 0.46 \% for the primary and secondary respectively) 
or major semi-axis ($\sim$0.2 \% both for the apparent and absolute values). 
The level of precision in $a$ also proves that the quality of the astrometric 
solution is very high. The 299 visibility measurements used provide good 
orbital phase coverage. The apparent and physical values of 
major semi-axis allow us to determine the parallax, thus the distance 
to the system, with a precision also close to 0.2 \%. Our value of the 
parallax -- 60.43(12) mas -- is in a relatively good agreement but almost 6
times more precise than 61.15(68) mas from the new reduction of the 
\emph{Hipparcos} data \citep{vLe07}. We were also able to precisely derive 
the magnitude difference in the $K$ band (282 $V^2$ measurements) but the 
accuracy for the $H$ band is much lower due to a lower number of $V^2$ 
measurements in $H$ (only 17). 

Our final error in the absolute masses of the BY Dra components is however much
higher -- 3.3 \% for both the primary and secondary. This is mainly due to
the inclination of the orbit of 154.4 deg. For such configurations, far from edge-on,
a small error in the angle propagates to a large error in the masses. Still
it is a considerably more accurate measurement compared to \citet{bod01}
of respectively 23\% and 25\% for the primary and secondary. This is
possible thanks to our superior RV data set (rms of $\sim$0.15 km s$^{-1}$ 
vs 2.3 km s$^{-1}$) and a longer time span of the astrometric $V^2$ data.

\section{Discussion}

\begin{figure*}
\includegraphics[width=\textwidth]{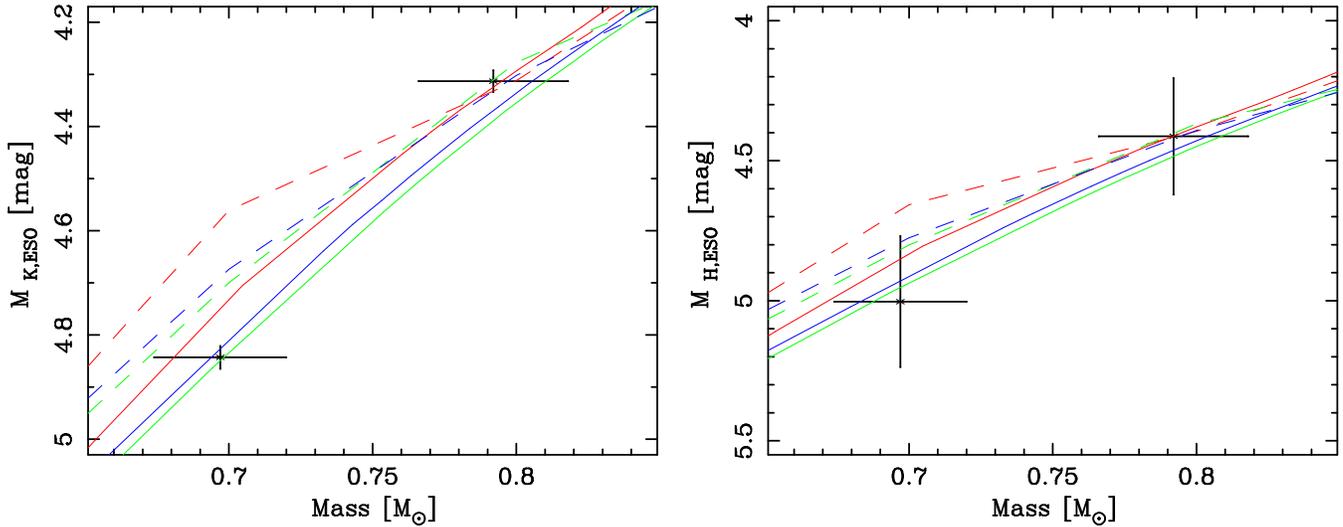}
\caption{Comparison of our results with the Yonsei-Yale isochrones in
the mass/$K$-band (left) and mass/$H$-band (right) absolute magnitude (in ESO system).
The isochrones for 1 Gyr are depicted with solid and for 60 Myr with dashed
lines. The isochrones for $Z=0.02$ are depicted with red, for 0.04 with green
and for 0.06 with blue lines. Colour version of the Figure is available 
in the on-line version of the article.}\label{fig_iso}
\end{figure*}

\subsection{Age and metallicity}
As it is pointed out by \citet{tor09}, the mass uncertainty
should be below 3 \% to be useful to perform 
reliable tests of the stellar evolution models. Our precision is close to
that but to go below 3 \% we would require a higher number of precise 
RVs or more $V^2$ measurements. Nevertheless, with our measurements we still 
can place some constraints on the evolutionary properties of BY Dra.
We focused on the age estimation to confirm or exclude the 
pre-main-sequence nature of the system. 
 
We compared our results with the Yonsei-Yale isochrones 
\citep[Y$^2$;][]{yi01,dem04}. We used our estimations of the magnitude 
differences, parallax and the apparent $H$ and $K$ magnitude from 2MASS
\citep{cut03} to derive the absolute $H$ and $K$ magnitudes of each 
component separately. Using the transformation equations from 
\citet[][with updates]{car01}\footnote{http://www.astro.caltech.edu/$\sim$jmc/2mass/v3/transformations/} 
we transformed them to the ESO photometric system \citep{vBl96} ,
in which the Y$^2$ isochrones are available. In Figure \ref{fig_iso}
we show our measurements in the mass/$K$-band (left) and mass/$H$-band
(right) absolute magnitude diagrams. 
For comparison we plot the isochrones for ages of
60~Myr (dashed) and 1~Gyr (solid lines) for three values of the metallicity: 
$Z$ = 0.02 (red), 0.04 (green) and 0.06 (blue). One can see that
the properties of the secondary component are not reproduced by the 60~Myr
isochrones, which means that it has already settled down on the main sequence. 
The formally best match is found for $t= 1$ Gyr and $Z=0.04$. 
No match was found for ages below 60 Myr for any $Z$ value 
nor for any age value for $Z<0.02$. For $t>5$ Gyr only isochrones with
metallicities higher than solar reproduce the data points. 
We thus conclude that BY~Dra is probably
between 0.2 and 5 Gyr old and is more metal-rich than the Sun. The
most probable values of age and $Z$ are 1-2 Gyr and 0.04 respectively
These facts make the pre-main-sequence scenario less probable. 

One should notice that the error bars in the masses are enlarged mainly
by the uncertainty in the inclination. Any change in $i$ would shift both 
components in the same direction -- towards higher or lower masses 
which would definitely not improve the fit.
We also have a large uncertainty in $M_H$, induced by the error
in $\Delta H$, which is so large due to a small number of $V^2$ measurements
in this band. Reduction of this uncertainty would allow for putting even
more stringent constrains on the nature of the system.
At the same time, $\Delta K$ 
is very well constrained and shows that the mass ratio 
$q\sim0.88$ is not inconsistent with the observed flux ratio, at 
least in the K band. Using the Y$^2$ set of isochrones we can 
estimate that the expected theoretical magnitude difference in $V$
for the stars having 0.792 and 0.697 M$_\odot$, should be close to 
0.9 mag. This is not in agreement with $1.15\pm0.1$ mag 
predicted by \citet{luc80}. However given even $\sim$0.2  
mag variation from spots \citep{chu66,pet92}, we can conclude that such a 
difference in V is possible for BY Dra even if 
it is a main-sequence system. 

To put additional constrains on the system's age, we further 
calculated the galactic space velocities $U,V,W$\footnote{Positive 
values of $U$, $V$ and $W$ indicate velocities toward the Galactic center, 
direction of rotation and north pole respectively} relatively to the 
local standard of rest \citep[LSR;][]{joh87}. We applied our values of 
radial systemic velocity and distance estimation together with
proper motion of $\mu_{\alpha}=185.92$ mas~yr$^{-1}$ and 
$\mu_{\delta}=-324.81$ mas~yr$^{-1}$ from the PPMX catalogue \citep{ros08}. 
Values of $U=28.2 \pm 0.1$, $V = -13.16 \pm 0.06$ and 
$W = -21.75 \pm 0.10$~km~s$^{-1}$ put BY~Dra outside of any known young moving 
group or group candidate \citep{zha09}, and at the transition area between the thin and 
thick galactic disk \citep{ben03,nor04}. This supports the possibility
of BY~Dra being not a PMS system.

\subsection{Spin-orbit (mis)alignment}

\begin{figure}
\includegraphics[width=\columnwidth]{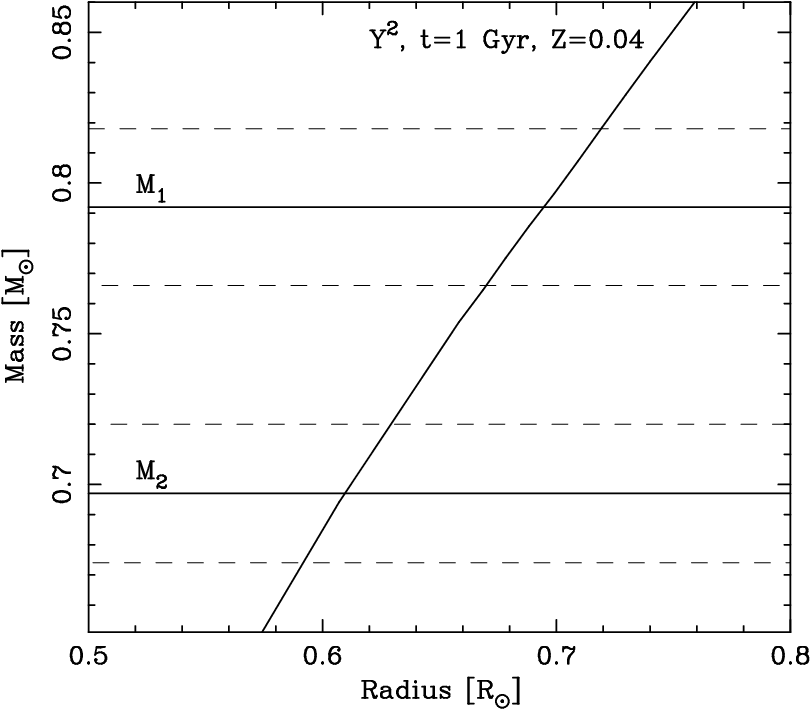}
\caption{Radii of the components as predicted from our estimations of 
BY Dra's parameters and the Y$^2$ best-matching isochrone. Horizontal 
solid lines represent our estimations of the masses and the dashed
lines the 1-sigma ranges of their respective uncertainties.}\label{fig_rpred}
\end{figure}

Using the masses and isochrones, we can estimate the radii
of each component of BY Dra. In Figure \ref{fig_rpred} we plot 
the Y$^2$ isochrone for 1 Gyr and $Z=0.4$ in a radius/mass plane.
As solid horizontal lines we plot the masses together with
their uncertainties (dashed lines). Other probable isochrones are very 
close to the chosen one and do not change the results of our analysis
significantly.

Our results predict values of $R_1=0.695\pm0.025$ and $R_2=0.61\pm0.02$
R$_\odot$. We can use this together with the rotational velocities and 
rotational period values from the literature to estimate the orbital
inclination angles. For this purpose we use $v_{rot,1}\sin i_{rot,1}
= 8.05 \pm 0.33$ and $v_{rot,2}\sin i_{rot,2}= 7.42 \pm 1.06$ km s$^{-1}$
from \citet{luc80} and $P_{rot}=3.8285$ d from \citet{pet92}. 
This implies $R_1\sin i_{rot,1}=0.61\pm0.03$ and 
$R_2\sin i_{rot,2}=0.56\pm0.08$. Most of the
authors attribute spots to the primary component and refer the 
$\sim$3.8 d period to its rotation. If so, from the values above we 
can estimate the rotational inclination of the primary to be 
$i_{rot,1}=61^{+12}_{-7}$ or $119^{+7}_{-12}$ deg. If we assume the 
secondary to rotate with the given period, we end up with
$i_{rot,2}=67$ or 113 deg and its uncertainties ranging from 50 to
130 deg. This means that $i_{rot,2}=90$ deg is also possible.
Our results are in a good agreement with values given by 
\citet{gle95} who derived $i_{rot,1}=60^{+11}_{-9}$ and 
$i_{rot,2}=85^{+5}_{-15}$ deg. None of the values however agrees 
with the orbital inclination $i_{orb}=154.4$ deg. This 
indicates the spin-orbit misalignment in the BY Dra system. 
Even if we consider that the theoretical radii are underestimated 
by about 15\%, a well known issue for late-type stars, we are not 
able to reproduce the observed $i_{orb}$. 

In the case of a spin-orbit alignment, and the literature
values of $v_{rot}\sin i_{rot}$, the radii would have to be $R_1\simeq1.41$ 
and $R_2\simeq1.30$ R$_\odot$. This would occur if the system was $\sim$3-4 Myr 
old, depending on the metallicity. In such a case both stars should be much 
brighter in the infrared than it is observed. The predicted $K$-magnitudes
difference for the two stars would be $\sim$0.2 and not
0.53 mag which is observed. One also would expect the
system to be a member of a young cluster, containing leftovers of the
primodial gas, but this is not observed as well.

However, the spin-orbit alignment should be observed for such a close
pair of $\sim$1 Gyr old stars \citep{hut81}. The source of the
discrepancy could be for example an overestimated rotational velocity.
\citet{gle95} in their analysis adopted a value of 3.6~km~s$^{-1}$,
given by \citet{str93}\footnote{In fact, in the current version of their
catalogue, \citet{str93} cite $v_{rot}$ values from \citet{luc80}.}. 
For this value of $v_{rot}\sin i_{rot}$ we get
$i_{rot,1} \simeq 157$~deg, which is very close to the observed
orbital inclination. Hence the spin-orbit alignment
may in fact be present in the BY~Dra system if \citet{luc80} have
overestimated their rotational velocity measurements by adopting
a too small macroturbulence velocity.  

Finally, let us note that a spin-orbit misalignment could manifest itself
through its impact on the apsidal precession rate \citep[for a review
see][]{maz08}. Unfortunately since BY Dra is not an eclipsing system
and our RVs and $V^2$s are not sufficiently accurate, a measurement of the
apsidal motion cannot be carried out. We attempted to fit for $\dot\omega$
but obtained statistically insignificant value.

\subsection{Rotation pseudo-synchronization}
In the case of eccentric orbits one can find a rotational period for which
an equilibrium is achieved. This 
equilibrium, called the \emph{pseudo-synchronization}, occurs when
the ratio of the orbital to the rotational period is:
\begin{equation}
\frac{P_{orb}}{P_{rot,ps}} = \frac{1 + 7.5e^2 + 5.625e^4 + 0.3125e^6}{(1-e^2)^{3/2}(1 + 3e^2 + 0.375e^4)}
\end{equation}
\citep{hut81,maz08}. For the observed eccentricity of BY~Dra we get
$P_{orb}/P_{rot,ps} = 1.559(3)$ or $P_{rot,ps}=3.833(8)$~d, which is in
a good agreement with the observed $P_{rot}=3.8285$~d, 
and $P_{orb}/P_{rot} = 1.561$. One may thus conclude that BY~Dra is 
in a rotational equilibrium. The predicted time scale of the 
pseudo-synchronization is in the case of BY~Dra similar but slightly 
shorter than for the spin-orbit alignment \citep{hut81}. Using the 
approximate formula for the synchronization time scale of late-type
stars given by \citet{dev08}, we get the value of the order of 10~Myr,
so still smaller than the age of the system. The above equation
was however derived for binaries with no additional companions
(see below) and the tidal evolution of BY Dra AB might be different if
the gravitational influence of the third body is taken into
account.

\subsection{Eccentricity and multiplicity of BY Dra}

The eccentricity of BY Dra AB ($e=0.3$) appears to be unusually high.
According to \cite{zahb89} a circularization of the orbit of a late type 
system such as BY~Dra should occur during the pre-main-sequence phase. 
This was one of the arguments for the PMS nature of BY~Dra. 
Based on \citet{zahb89} we can estimate that in the case of BY~Dra 
the eccentricity should drop to a few percent over $\sim 10^5$ years.
 
However, BY~Dra is a hierarchical triple system, with a 
distant common proper motion companion. The projected separation of 
16.7 arc-sec and our distance determination indicate a projected physical 
separation of 277 AU. As estimated by \citet{zuc97} its mass is about 
0.13 M$_\odot$ and assuming a circular orbit, it corresponds to an 
orbital period of about 2050 years. It is conceivable that the observed
eccentricity of the BY~Dra~AB pair could be explained by the presence 
of the companion through the Mazeh-Shaham mechanism 
which results in a cyclic eccentricity variation, 
known as the Kozai cycles \citep{koz62,maz79,fab07,maz08}.

Let us denote all the orbital and physical parameters of an
unknown perturber by the index $X$.
In order to put some constrains on the properties of the perturber which
would induce sufficiently strong Kozai cycles, we followed the analysis 
of \citet{fab07}. The two main conditions which have to be met 
in order to produce the observed eccentricity are: (1) a sufficiently
large relative inclination $i_{rel}$ of the binary (inner) and the perturber's 
(outer) orbit; (2) the Kozai cycles time-scale $\tau$ must be shorter than 
the period of the inner orbit's pericenter precession. For the BY~Dra~AB 
pair the relativistic precession is the dominant one, being at least
an order of magnitude faster than any other \citep[tidal or 
rotational;][]{fab07}. The precession period is $U_{GR} \simeq 25000$~yr,
which corresponds to $\dot{\omega}_{GR} = 8.0 \times 10^{-12}$~rad~s$^{-1}$. 
Using the formalism of \citet{fab07} we can estimate that
the observed eccentricity of BY~Dra can be induced when 
$\tau \dot{\omega}_{GR}|_{e=0} \le 2.796$ (in SI units),
where the term $\dot{\omega}_{GR}$ is computed for $e=0$. 
From this we can derive the following conditions for the
parameters of the perturbing body:
\begin{equation}\label{rel_koz_p}
P^2_X \frac{M_1+M_2+M_X}{M_X} \left( 1-e^2_X \right)^{3/2}<938.21
\end{equation}
or
\begin{equation}\label{rel_koz_a}
\frac{a^3_X}{M_X} \left( 1-e^2_X \right)^{3/2}<938.21,
\end{equation}
where the orbital period is given in years, major semi-axis in AU
and all masses in M$_\odot$. 
The condition $\tau \dot{\omega}_{GR}|_{e=0} \le 2.796$ 
also allows us to deduce that the relative inclination of 
the two orbits must be larger than $78^\circ$ (or smaller than $102^\circ$). 
This value can be confirmed by the results of \citet{for00}, 
which for the mass ratio of BY~Dra~AB predict $i_{rel} \ga 75^\circ$. 
The relatively narrow range of $i_{rel}$ allows us to put some 
usefull constrains (which include possible values of the longitudes
of ascending nodes) on the {\it absolute} value of the perturber's 
orbital inclination: $i_X \in[52.4^\circ, 127.6^\circ]$.

\begin{figure}
\includegraphics[width=\columnwidth]{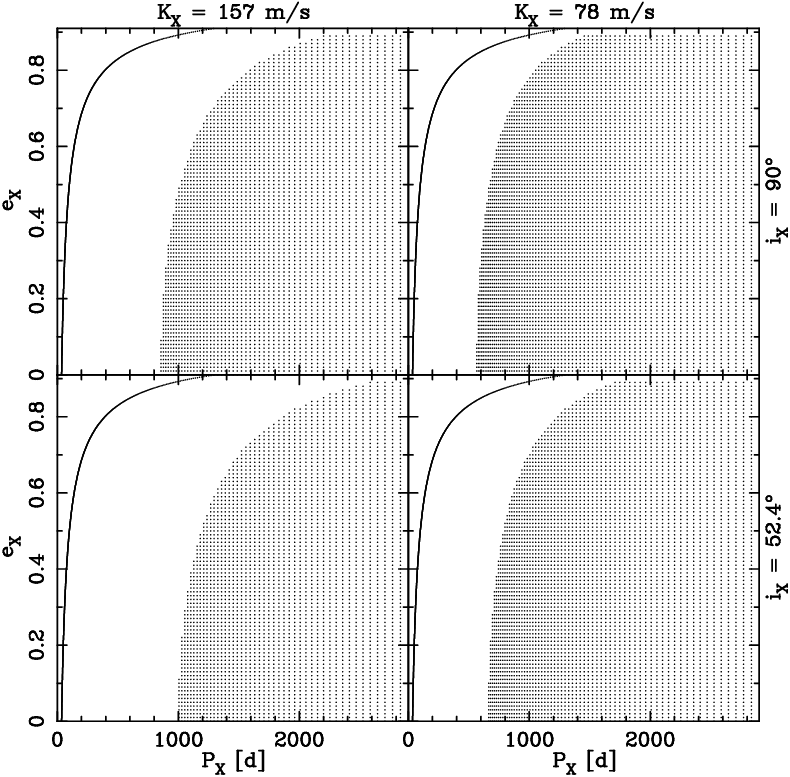}
\caption{Conditions necessary to reproduce the observed eccentricity
of BY~Dra through the Mazeh-Shaham mechanism by bodies which would
produce the RV signal with the semi-amplitude of 157 and 78
m s$^{-1}$ (the lower $rms$ of the RV orbital solution and half of its
value, left and right column respectively), and having orbital 
inclinations of 90 (top row) and 52.4$^\circ$ (bottom row).
The shaded areas refer to the periods and eccentricities which
would not allow to induce sufficiently strong Kozai cycles.
The lower limit of the period is 35 d (the semimajor axis of 0.24 AU) 
which refers to the shortest stable {\it circular} orbit \citep{hol99}. 
The area above the solid line corresponds to the eccentric orbits
whose periastron distance is within the instability zone
(i.e. is shorter than 0.24 AU). The upper limit in the eccentricities
shown in the figures is 0.9. 
}\label{fig_kozai}
\end{figure}

\begin{figure}
\includegraphics[width=\columnwidth]{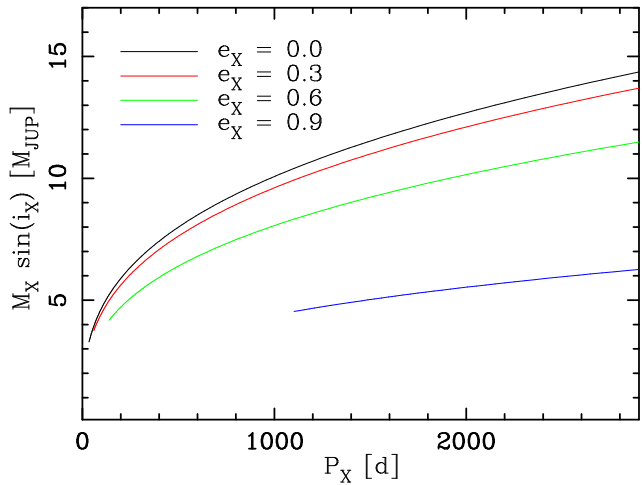}
\caption{Limits for $M_X \sin(i_X)$ for a putative
fourth body estimated from our RV measurements, for several 
values of eccentricities. The value of the $rms=157$ m s$^{-1}$ was assumed
to be the RV semi-amplitude. The lower limit of the period is 35 d 
(the semimajor axis of 0.24 AU) which refers to the shortest stable 
{\it circular} orbit \citep{hol99}. 
For eccentric orbits the limits are terminated at the shortest
periods having the distance of periastron larger than 0.24 AU
(around 60, 139 and 1107 days). Using $i_X = 52.4^\circ$ increases
the detection limits by about 26\% ($1./\sin(52.4^\circ)\simeq 1.26$). 
Colour version of the Figure is available in the
on-line version of the manuscript.
}\label{fig_detect}
\end{figure}

In Figure \ref{fig_kozai} we present results of our analysis. 
We use relation \ref{rel_koz_p} to check whether a body of a 
given orbital properties (i.e. period and eccentricity) can
produce sufficient Kozai cycles. For the mass $M_X$ we have 
taken mass of a body that in an orbit of a given $P_X$, $e_X$, 
and $i_X = 52.4$ or 90$^\circ$ would produce an RV modulation of 
the inner pair at the level of the smaller $rms$ (157 m s$^{-1}$) or
half of that. The four panels show the $P_X/e_X$ parameter space for 
the two values of inclinations and RV semi-amplitudes. The shaded 
areas correspond to the values of $P_X$ and $e_X$ for which the 
observed eccentricity of BY~Dra~AB would not be induced. 
The solid line shows the short-period stability border, calculated 
in such way that for a given eccentricity the orbit has its 
periastron at $\sim$0.24 AU which refers to the smallest stable
{\it circular} orbit \citep[with $P\simeq35$ d;][]{hol99}. 
The long-period cut-off at
2900 days is mainly for the clarity, but it is close to double
time span of our observations (1513~d) which means that RV modulations
with periods around 3000 d could in principle be detected.
The corresponding fourth body detection limits in terms 
of $M_X \sin(i_X)$ as a function of its orbital
period and eccentricity are presented in the Figure \ref{fig_detect}.

From those two figures one can deduce that if the observed eccentricity
is an effect of the Mazeh-Shaham mechanism, the perturber should be
in an orbit of up to single years (major semi axes from $\sim 0.2$ 
to $\sim 2$ AU) and have its mass in the planetary regime. 
This is consistent with the fact that the interferometric $V^2$
measurements are fully-consistent with a two-disc model, so no 
additional light is detected. It is not however 
excluded that the eccentricity of the perturber is very large which would 
allow more massive bodies in long period orbits to induce the 
Kozai cycles and remain undetectable by the RVs.
The component C discovered by \citet{zuc97} could be the perturbing
body but with the relation \ref{rel_koz_a} it seems that it is
not very likely.  
For the estimated mass $M_3 = 0.13$ M$_\odot$ the orbital
eccentricity $e_3$ would have to be larger than 0.98, assuming 
a fortunate but improbable case that the star is currently seen
exactly at the apocenter, and $\Omega_3=0^\circ$, 
thus $a_3$ [AU] $= 277/(1+e_3)$. For less fortunate cases the
value of $e_3$ would have to be even larger.

The other possibility is the presence of a putative fourth body
reported in the \emph{Hipparcos} catalog.
\citet{bod01} inspected the available RV data in order to find a 
114 day period predicted by the \emph{Hipparcos} catalog and with a
high level of confidence they excluded the existence of such a 
period in the spectroscopic data. With the relation \ref{rel_koz_p}
we can show that if a body in a $P_H=114.02$ d circular orbit exists, 
it would have to have $M_H \ga 0.15$ M$_{JUP}$. From the Figure
\ref{fig_detect} we can put an upper mass limit of 5.44 M$_{JUP}$,
taking into account the reported inclination of 113.21$^\circ$,
which itself is within the allowed limits. The reported
major semi-axis of the \emph{Hipparcos} photocentric orbital solution is
$a_{12,H} = 0.0515$ AU (at the distance to BY~Dra). Assuming the
maximum mass ratio $M_H/(M_1 + M_2)=0.0035$, the 4-th body
barycentric major semi-axis would be 16.07 AU (after correcting
for the inclination), but the Kepler's 3'rd law predicts the
major semi-axis of 0.53 AU for the given period and masses.
Such a body would produce the RV signal much stronger than 157 m s$^{-1}$.
We can thus conclude that the \emph{Hipparcos} solution is
unrealistic.

\section{Summary}
We present the most precise orbital and physical parameters of an
important astrophysical object -- the low-mass SB2 BY~Draconis, a 
prototype of an entire class of variable stars. We reach a level of 
precision which allows us to put important constrains on the nature of this
object. We conclude that this is not a pre-main-sequence system, 
despite its high orbital eccentricity and a possible spin-orbit 
misalignment. However, the gravitational influence of a 4-th,
yet undetected body in the system may explain the observed value 
of $e$ and the spin-orbit alignment may be inferred from the available 
data if a smaller than claimed in the literature value of the rotational 
velocity is used. The observed rotational 
period and the eccentricity suggest that the BY Dra AB system is in the 
rotational equilibrium. However, if the observed eccentricity is indeed 
due to the presence of the 4-th body, the putatiuve companion may have 
its mass in the planetary regime. The whole dynamical and tidal picture 
of this system is more complicated than we previoulsy thought 
and deserving perhaps a dedicated theoretical, observational
and numerical analysis.

\section*{Acknowledgments}
We would like to thank Prof. Tsevi Mazeh for his {invaluable} comments
and suggestions, and Arne Rau for carrying out the Keck I/HIRES 
observations in the years 2006-2007. The authors wish to recognize 
and acknowledge the very significant cultural role and reverence that 
the summit of Mauna Kea has always had within the indigenous Hawaiian
community. We are most fortunate to have the opportunity
to conduct observations from this mountain.
This work benefits from the efforts of the PTI collaboration members who have
each contributed to the development of an extremely reliable observational
instrument. We thank PTI's night assistant Kevin
Rykoski for his efforts to maintain PTI in excellent condition and operating
PTI. 

This research was co-financed by the European Social Fund and the national
budget of the Republic of Poland within the framework of the Integrated
Regional Operational Programme, Measure 2.6. Regional innovation
strategies and transfer of knowledge - an individual project of the
Kuyavian-Pomeranian Voivodship ``Scholarships for Ph.D. students
2008/2009 - IROP'', and by the grant N N203 379936 from the Ministry of 
Science and Higher Education. Support for K.G.H. is provided by Centro 
de Astrof\'{i}sica FONDAP Proyecto 15010003, M.K. is supported by the Foundation for 
Polish Science through a FOCUS grant and fellowship and by the Polish 
Ministry of Science and Higher Education through grants N N203 005 32/0449
and N N203302035.
M.W.M. acknowledges support from the Townes Fellowship Program, an internal 
UC Berkeley SSL grant, and the State of Tennessee Center of Excellence program. 
This research was supported in part by the National Science Foundation 
under Grant No PHY05-51164. The observations on the
TNG/SARG have been funded by the Optical Infrared Coordination network (OPTICON),
a major international collaboration supported by the Research Infrastructures 
Programme of the European Commissions Sixth Framework Programme.

This research has made use of the Simbad database, operated at CDS, Strasbourg, 
France, and of data products from the Two Micron All Sky
Survey, which is a joint project of the University of Massachusetts and the
Infrared Processing and Analysis Center/California Institute of Technology,
funded by the National Aeronautics and Space Administration and the National
Science Foundation.

\appendix

\bsp

\label{lastpage}


\begin{thebibliography}{99}
\bibitem[\protect\citeauthoryear{Bagnoulo \& Gies}{1991}]{ban91}
	Bagnoulo W.G., Jr., Gies D.R., 1991, ApJ, 376, 266
\bibitem[\protect\citeauthoryear{Barnes, Evans \& Moffet}{1978}]{bar78}
	Barnes E.S., Evans D.S., Moffet T.J., 1978, MNRAS, 183, 285
\bibitem[\protect\citeauthoryear{Bensby et al.}{2003}]{ben03} 
	Bensby T., Feltzing S., Lundstr\"om I., 2003, A\&A, 410, 527
\bibitem[\protect\citeauthoryear{Boden}{1999}]{bod99}
	Boden A.F., in Lawson P.R., ed., Principles of Long Baseline
	Stellar Interferometry, JPL Publication, Pasadena, p.9
\bibitem[\protect\citeauthoryear{Boden \& Lane}{2001}]{bod01}
	Boden A.F., Lane B.F., 2001, ApJ, 547, 1071 
\bibitem[\protect\citeauthoryear{Bopp \& Evans}{1973}]{bop73}
	Bopp B.W., Evans D.S., 1973, MNRAS, 164, 343
\bibitem[\protect\citeauthoryear{Bopp, Noah \& Klimke}{1980}]{bop80}
	Bopp B.W., Noah P., Klimke A., 1980, AJ, 85, 1386
\bibitem[\protect\citeauthoryear{Carpenter}{2001}]{car01}
	Carpenter J.M., 2001, AJ, 121, 2851 
\bibitem[\protect\citeauthoryear{Chugainov}{1966}]{chu66}
	Chugainov P.S., 1966, IBVS, 122, 1
\bibitem[\protect\citeauthoryear{Colavita et al.}{1999}]{col99}
	Colavita M.M., 1997, ApJ, 510, 505
\bibitem[\protect\citeauthoryear{Cristaldi \& Rodono}{1968}]{cri68}
	Cristaldi S., Rodono M., 1968, IBVS, 252, 2
\bibitem[\protect\citeauthoryear{Cristaldi \& Rodono}{1970}]{cri70}
	Cristaldi S., Rodono M., 1970, A\&AS, 2, 223
\bibitem[\protect\citeauthoryear{Cristaldi \& Rodono}{1971}]{cri71}
	Cristaldi S., Rodono M., 1971, A\&A, 12, 152
\bibitem[\protect\citeauthoryear{Cutri et al.}{2003}]{cut03}
	Cutri R.M., et al., 2003, The IRSA 2MASS All-Sky Catalog of Point Sources, 
	NASA/IPAC Infrared Science Archive
\bibitem[\protect\citeauthoryear{Demarque et al.}{2004}]{dem04}
	Demarque P., Woo J.-H., Kim Y.-C., Yi S.K., 2004, ApJS, 155, 667
\bibitem[\protect\citeauthoryear{Devor et al.}{2008}]{dev08}
	Devor J., et al., 2008, ApJ, 687, 1253
\bibitem[\protect\citeauthoryear{ESA}{1997}]{esa97}
	ESA, 1997, The Hipparcos and Tycho Catalogues, ESA SP-1200 
\bibitem[\protect\citeauthoryear{Fabrycky \& Tremaine}{2007}]{fab07} 
	Fabrycky D., Tremaine S., 2007, ApJ, 669, 1298 
\bibitem[\protect\citeauthoryear{Ford, Kozinsky \& Rasio}{2000}]{for00}
	Ford E., Kozinsky B., Rasio F.A., 2000, ApJ, 535, 385
\bibitem[\protect\citeauthoryear{G\l\c{e}bocki \& Stawikowski}{1995}]{gle95}
	G\l\c{e}bocki R., Stawikowski A., 1995, AcA, 45, 725
\bibitem[\protect\citeauthoryear{He\l miniak \& Konacki}{2011}]{hel11a}
        He\l miniak K.G., Konacki M., 2011, A\&A, 526, A29
\bibitem[\protect\citeauthoryear{He\l miniak et al.}{2011}]{hel11b}
        He\l miniak K.G., Konacki M., Z\l oczewski K., et al. 2011, A\&A, 527, A14
\bibitem[\protect\citeauthoryear{Holman \& Wiegert}{1999}]{hol99}
	Holman M. J., Wiegert P. A., 1999, AJ, 117, 621
\bibitem[\protect\citeauthoryear{Hut}{1981}]{hut81}
	Hut P., 1981, A\&A, 99, 126
\bibitem[\protect\citeauthoryear{Johnson \& Soderblom}{1987}]{joh87}
	Johnson D.R.H., Soderblom D.R., 1987,	AJ, 93, 864
\bibitem[\protect\citeauthoryear{Krzemi\'{n}ski}{1969}]{krz69}
	Krzemi\'{n}ski W., 1969, in Kunar S., ed., Low Luminosity Stars, 
	Gordon and Breach Publishing Co., London, p.57
\bibitem[\protect\citeauthoryear{Konacki}{2009}]{kon09}
	Konacki M., 2009, IAU Symposium, 253, 141
\bibitem[\protect\citeauthoryear{Konacki et al.}{2009}]{kon09a}
	Konacki M., Muterspaugh M.W., Kulkarni S.R., He{\l}miniak K.G., 
	2009, ApJ, 704, 513
\bibitem[\protect\citeauthoryear{Konacki et al.}{2010}]{kon10}
	Konacki M., Muterspaugh M.W., Kulkarni S.R., He{\l}miniak K.G.,
	2010, ApJ, 719, 1293
\bibitem[\protect\citeauthoryear{Kozai}{1962}]{koz62}
	Kozai Y., 1962, AJ, 67, 591
\bibitem[\protect\citeauthoryear{Krzemi\'{n}ski \& Kraft}{1967}]{krz67}
	Krzemi\'{n}ski W., Kraft R.P., 1967, AJ, 72, 307
\bibitem[\protect\citeauthoryear{Kurucz}{1995}]{kur95}
	Kurucz R.L., ASP Conf. Ser. 78: Astrophysical Applications of 
	Powerfull New Databases, 205
\bibitem[\protect\citeauthoryear{Landin, Mendes \& Vaz}{2009}]{lan09}
	Landin N.R., Mendes L.T.S., Vaz L. P. R., 2009, A\&A, 494, 209
\bibitem[\protect\citeauthoryear{Lucke \& Mayor}{1980}]{luc80}
	Lucke P.B., Mayor M., 1980, A\&A, 92, 182
\bibitem[\protect\citeauthoryear{Masani et al.}{1955}]{mas55}
	Masani A., Broglia P., Pestarion E., 1955, Mem. Soc. Astr. Ital., 26, 183
\bibitem[\protect\citeauthoryear{Mazeh}{2008}]{maz08} 
	Mazeh T., 2008, EAS, 29, 1 
\bibitem[\protect\citeauthoryear{Mazeh \& Shaham}{1979}]{maz79}
	Mazeh T., Shaham J., 1979, A\&A, 77, 145
\bibitem[\protect\citeauthoryear{M\"{u}nch}{1944}]{mun44} 
	M\"{u}nch G., 1944, ApJ, 99, 271
\bibitem[\protect\citeauthoryear{Nordstr\"om et al.}{2004}]{nor04}
	Nordstr\"om B., et al. 2004, A\&A, 418, 989
\bibitem[\protect\citeauthoryear{Pettersen, Olah \& Sandmann}{1992}]{pet92}
	Pettersen B.R., Olah K., Sandmann W.H., 1992, A\&AS, 96, 497
\bibitem[\protect\citeauthoryear{Popper}{1953}]{pop53} 
	Popper D.M., 1953, PASP, 65, 278
\bibitem[\protect\citeauthoryear{R\"oser et al.}{2008}]{ros08}
	R\"oser S., Schilbach E., Schwan H., Kharchenko N.V., Piskunov A.E., Scholz R.-D., 2008, A\$A, 488, 401
\bibitem[\protect\citeauthoryear{Strassmeier et al.}{1993}]{str93}
	Strassmeier K.G., Hall D.S., Fekel F.C., Scheck M., 1993, A\&AS, 100, 173
\bibitem[\protect\citeauthoryear{Torres, Andersen \& Gim\'{e}nez}{2010}]{tor09}
	Torres G., Andersen J. \& Gimen\'ez A., 2010, A\&A Rev, 18, 67
\bibitem[\protect\citeauthoryear{van der Bliek, Manfroid \& Bouchet}{1996}]{vBl96}
	van der Bliek N.S., Manfroid J., Bouchet P., 1996, A\&AS, 119, 547
\bibitem[\protect\citeauthoryear{van de Kamp}{1967}]{kam67}
	van de Kamp P., 1967, Principles of Astrometry, Freeman, San Francisco
\bibitem[\protect\citeauthoryear{van Leeuwen}{2007}]{vLe07}
	van Leeuwen F., 2007, A\&A, 474, 653
\bibitem[\protect\citeauthoryear{Vogt \& Fekel}{1979}]{vog79} 
	Vogt S.S., Fekel F., 1979, ApJ, 234, 958
\bibitem[\protect\citeauthoryear{Wilson \& Devinney}{1971}]{wil71}
	Wilson R.E., Devinney R.J., 1971, ApJ, 166, 605
\bibitem[\protect\citeauthoryear{Yi et al.}{2001}]{yi01}
	Yi S.K., Demarque P., Kim Y.-C., Lee Y.-W., Ree C.H., Lejeune T., Barnes S.,
	2001, ApJS, 136, 417
\bibitem[\protect\citeauthoryear{Zahn}{1977}]{zah77}
	Zahn J.-P., 1977, A\&A, 57, 383
\bibitem[\protect\citeauthoryear{Zahn}{1989}]{zah89}
	Zahn J.-P., 1989, A\&A, 220, 112
\bibitem[\protect\citeauthoryear{Zahn}{2008}]{zah08}
	Zahn J.-P., 2008, EAS, 29, 67
\bibitem[\protect\citeauthoryear{Zahn \& Bouchet}{1989}]{zahb89}
	Zahn J.-P., Bouchet L., 1989, A\&A, 223, 112
\bibitem[\protect\citeauthoryear{Zhao et al.}{2009}]{zha09}
	Zhao J., Zhao G., Chen Y., 2009, ApJ, 692, L113
\bibitem[\protect\citeauthoryear{Zucker \& Mazeh}{1994}]{zuc94}
	Zucker S., Mazeh T., 1994, ApJ, 420, 806
\bibitem[\protect\citeauthoryear{Zuckerman et al.}{1997}]{zuc97}
	Zuckerman B., Webb R.A., Becklin E.E., McLean I.S., Malkan M.A.,
	1997, AJ, 114, 805

\end{thebibliography}
\end{document}